# Generation and detection of spin current in iridate/manganite heterostructures

[1,2]Ulev G.D., [1]Ovsyannikov G.A., [1]Constantinian K.Y., [1,3]Shadrin A.V., [1]Moskal I.E., [1]Lega P.V.

[1]V.A. Kotel'nikov Institute of Radio Engineering and Electronics Russian Academy of Sciences. 125009, Moscow, Russia.
[2]National Research University Higher School of Economics, Faculty of Physics, 101000, Moscow, Russia
[3]Moscow Institute of Physics and Technology (National Research University), 141701, Moscow region, Dolgoprudny, Russia.

The present study focuses on experimental investigations of spin current across the interface of an iridate/manganite heterostructure ($SrIrO_3/La_{0.7}Sr_{0.3}MnO_3$) consisting of oxide epitaxial films with nanometer thickness. Pure spin current is induced by microwave irradiation in the GHz frequency band, specifically under conditions of ferromagnetic resonance. The detection of spin current is achieved through the inverse spin-Hall effect, which measures the charge current arising on the electrically conductive $SrIrO_3$ film with strong spin-orbit interaction. To quantify the efficiency of spin current conversion to charge current, the angular dependences of spin magnetoresistance of the iridate/manganite interface are measured, thereby determining the spin-Hall angle.

## 1. Introduction

Spintronic devices offer a potential solution to the energy dissipation issue in microelectronics, by utilizing spin transfer (spin current), which does not generate heat. However, the detection and generation of pure spin current (without charge transfer) necessitate a different approach compared to traditional charge-based electronic systems. One technique for generating pure spin current involves the precession of ferromagnetic magnetization induced by a microwave magnetic field under ferromagnetic resonance. The magnitude of the spin current depends on the amplitude of magnetization precession and the spin-mixing conductance of the interface ($g^{\uparrow\downarrow}$), which consists of real and imaginary components. The spin current can be detected through the inverse spin-Hall effect (ISHE) in a metal ($N_{so}$) with strong spin-orbit interaction, where the spin-Hall magnetoresistance (SMR) occurs in the ferromagnetic/normal (F/$N_{so}$) heterostructure. [1-6]. The measurement of angle dependences of SMR is a useful tool for determining the spin Hall angle $\theta_{SH}$, which characterizes the efficiency of converting spin current into charge current [6-8].

Previous studies have carried out spin current excitation at FMR and its using the ISHE technique in F/$N_{so}$ interfaces, [2, 3] with platinum (Pt) as the $N_{so}$ metal and permalloy (NiFe) as the F metal. Other investigations have explored structures with the $N_{so}$ metal deposited on top of the insulating ferromagnetic, iron-yttrium garnet (YIG)[9, 10]. Notably, metals with strong-orbital interaction (SOI) have been predominantly used [10, 11].

In this study, strontium iridate $SrIrO_3$. a 5d transition metal oxide [12, 13] known for its strong spin-orbital interaction and electron-electron, is utilized. The combination of these effects leads to non-trivial quantum phases [14] and enables control over magnetic anisotropy [5]. Previous studies have examined the charge-spin coupling in $SrIrO_3$ structures with a metallic ferromagnetic deposited on top: $SrIrO_3$/Py [15, 16], and $SrIrO_3$/$Co_{1-x}Tb_x$ [17], demonstrating the presence of an anomalously large spin-Hall angle due to the impact of spin-orbit interaction in $SrIrO_3$[15, 16].

The utilization of oxide materials enables the creation of heterostructures with atomically smooth interfaces when thin epitaxial films of $SrIrO_3$ are grown on epitaxial films of $La_{0.7}Sr_{0.3}MnO_3$[17-19] These films are typically deposited on substrates such as $(LaAlO_3)_{0.3}(Sr_2AlTaO_6)_{0.7}$ (LSAT), $NdGaO_3$, or $SrTiO_3$ using laser ablation [18-20], or magnetron sputtering at high temperature [12]. Previous studies have demonstrated that an increase in Hilbert damping in $SrIrO_3$/$La_{0.7}Sr_{0.3}MnO_3$ heterostructure is caused by the flow of spin current across the interface [12, 18, 19]. Furthermore, experimental investigations of spin current characteristics, under the influence of spin pumping, have revealed that the anisotropic magnetoresistance of $La_{0.7}Sr_{0.3}MnO_3$ also contributes to the overall response, along with the component generated by spin current [8, 12, 13, 18].

2. **Materials and Methods**.

The epitaxial films of $SrIrO_3$ (further, SIO3) and manganite $La_{0.7}Sr_{0.3}MnO_3$ (LSMO) with thicknesses ranging from 10 to 50 nm were grown on single crystal (110)$NdGaO_3$ substrates using radio frequency magnetron sputtering. The substrate temperatures during deposition ranged from 770 to 800˚C and a a mixture of Ar and $O_2$ gases was used at a total gas pressure of 0.3-0.5 mbar [12].

The crystal structure of resulting heterostructures was analyzed using X-ray diffraction and transmission electron microscopy (TEM). Figure 1 illustrates a TEM image of a cross-section of the heterostructure, show the SIO3/LSMO interface as well as the interface between the LSMO film and the $NdGaO_3$ substrate. The crystal lattice of SIO3 and LSMO is described as a distorted pseudo-cube with parameters $a_{SIO}$ = 0.396 nm and $a_{LSMO}$ = 0.389 nm correspondingly, as the heterostructure growth occurs through a cube-on-cube mechanism with the following ratios: $(001)SrIrO_3||(001)La_{0.7}Sr_{0.3}MnO_3||(110)NdGaO_3$, and $[100]SrIrO_3||[100]La_{0.7}Sr_{0.3}MnO_3||[001]NdGaO_3$ [12].

3. **Results and Discussion**.

In the SIO3/LSMO heterostructure, the paramagnetic SIO3 film acts as a normal metal with strong SOI, while the ferromagnetic LSMO is a magnetic half-metal with nearly 100% magnetic polarization at low temperatures.

In experiments involving spin pumping at microwaves, the essential parameters are the spin diffusion length, which characterizes the dissipation of spin current in the $N_{so}$ metal, the spin-Hall angle $\theta_{SH}$, which represents the ratio of spin to charge currents at the $F/N_{so}$ interface, and the spin mixing conductance $g^{\uparrow\downarrow}$, which is determined by the electron scattering matrix at $F/N_{so}$ and describes the transparency of angular magnetic momentum transfer through the interface [4, 5].

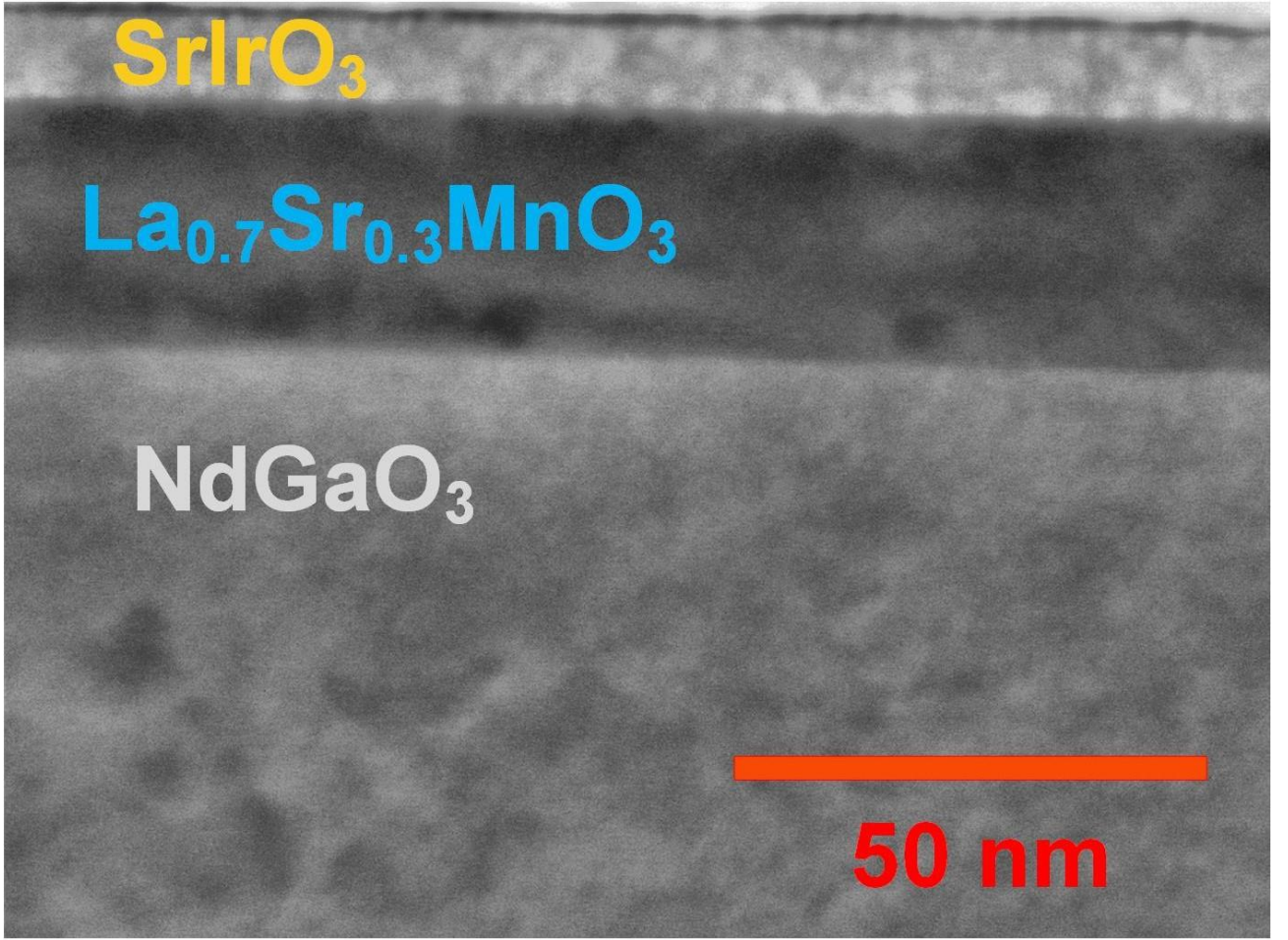

Fig. 1. Cross section of the SrIrO$_3$/La$_{0.7}$Sr$_{0.3}$MnO$_3$ heterostructure on an NdGaO$_3$ substrate, obtained through transmission electron microscope. The additional Pt platinum film, which was deposited for ion etching, was subsequently removed

**3.1 Spin current generation.**

During FMR spin pumping, a spin current $j_S$ flows through the SIO3/LSMO interface, characterized by the spin mixing conductance $g^{\uparrow\downarrow}$, which consists of both a real ( Re $g^{\uparrow\downarrow}$ ) and an imaginary part ( Im $g^{\uparrow\downarrow}$ ), as well by the amplitude of magnetic moment precession m induces by the magnetic component of the microwave field [4, 13]:

$$j_s = \frac{h}{4\pi}\left(\mathrm{Re}\,g^{\uparrow\downarrow}\, \mathrm{m}\frac{d\mathrm{m}}{dt} + \mathrm{Im}\,g^{\uparrow\downarrow}\frac{d\mathrm{m}}{dt}\right) \qquad (1)$$

The spin current was measured by observing the voltage across a sample shaped as a strip of the SIO3/LSMO heterostructure with metal Pt contacts. The microwave magnetic field was applied by placing the sample on a microstrip line, allowing measurements in the frequency range of f = 2–20 GHz [10]. A dc magnetic field (H) was set in the plane of the substrate and perpendicular to the emerging charge current (along the Y axis), while the microwave magnetic field was generated by a short-circuited microstrip line with the magnetic component directed along the X axis (see inset to Fig.2). Precession of the magnetization of the LSMO film induces a spin current perpendicular to the SIO3/LSMO interface (Z axis), which can be detected though voltage measurements using ISHE. The charge current $j_Q$ is related to the spin current $j_S$ through the spin Hall angle $\theta_{SH}$, a dimensionless parameter[2, 6]:

$$\vec{j}_Q = \theta_{SH} \frac{2e}{\hbar}\left[\vec{n} \times \vec{j}_S\right] \qquad (2)$$

where $\vec{n}$ is the unit vector of the spin momentum direction.

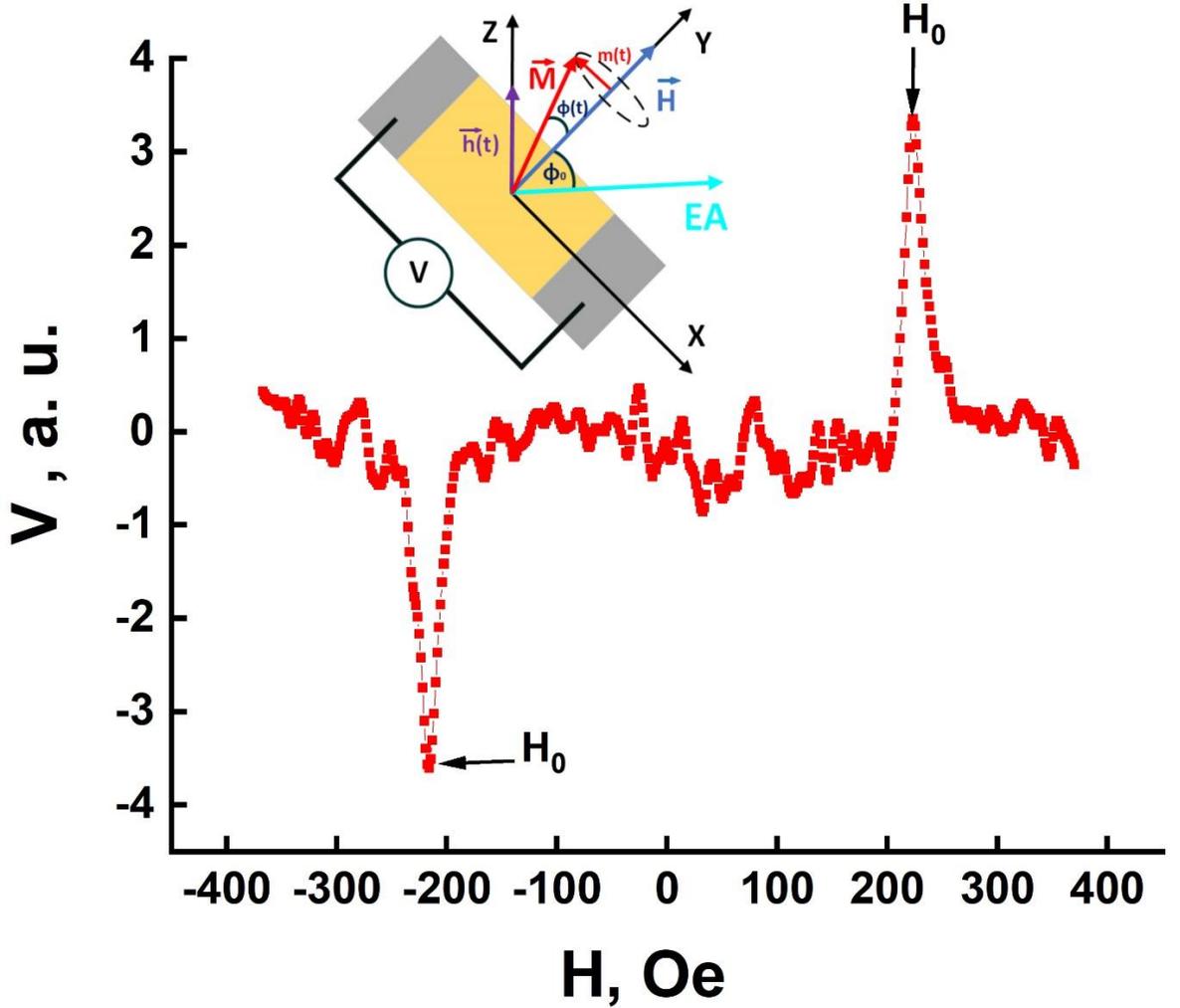

Fig. 2. The dependence of the response of the SrIrO$_3$/La$_{0.7}$Sr$_{0.3}$MnO$_3$ heterostructure on the dc magnetic field, V(H), when a microwave field is applied at a frequency of f = 2.3 GHz at a temperature of T = 300 K with a power of 30 mW. The inset on the right provides the topological representation of the heterostructure, indicating the directions of the microwave magnetic component (h(t)) and dc magnetic field (H), as well as the direction of the flow of the charge current (registered voltage V along the X axis) caused by the spin current.

Fig. 2 illustrates the dependence of the voltage V(H) on the dc magnetic field for the SIO3/LSMO heterostructure when subjected to f=2.3 GHz microwave field with power of 30 mW at T=300 K. The response V(H) undergoes a reversal in sign when the direction of the dc magnetic field is changed, indicating a change in the direction of the induced charge current.

This change is attributed to the reversal of the dc magnetic field direction (see eq.(2)). Additionally, experimental observations show some asymmetry in the response amplitudes for opposite directions of the H field, which is caused by microwave detection by Pt contacts. The anisotropic magnetoresistance (AMR) response also contributes to the observed asymmetry but does not depend on the direction of the dc magnetic field under certain conditions [8, 21].

Fig. 3 presents the voltage response caused by spin current at a frequency of f=2.3 GHz and T=300 K. This response is obtained from the half difference of peaks at opposite directions of the dc magnetic field (Fig. 2). The shape of the spin response is described by the Lorenz function. The contributions from microwave detection by Pt contacts and AMR are an order magnitude smaller than the amplitude of the spin voltage response.

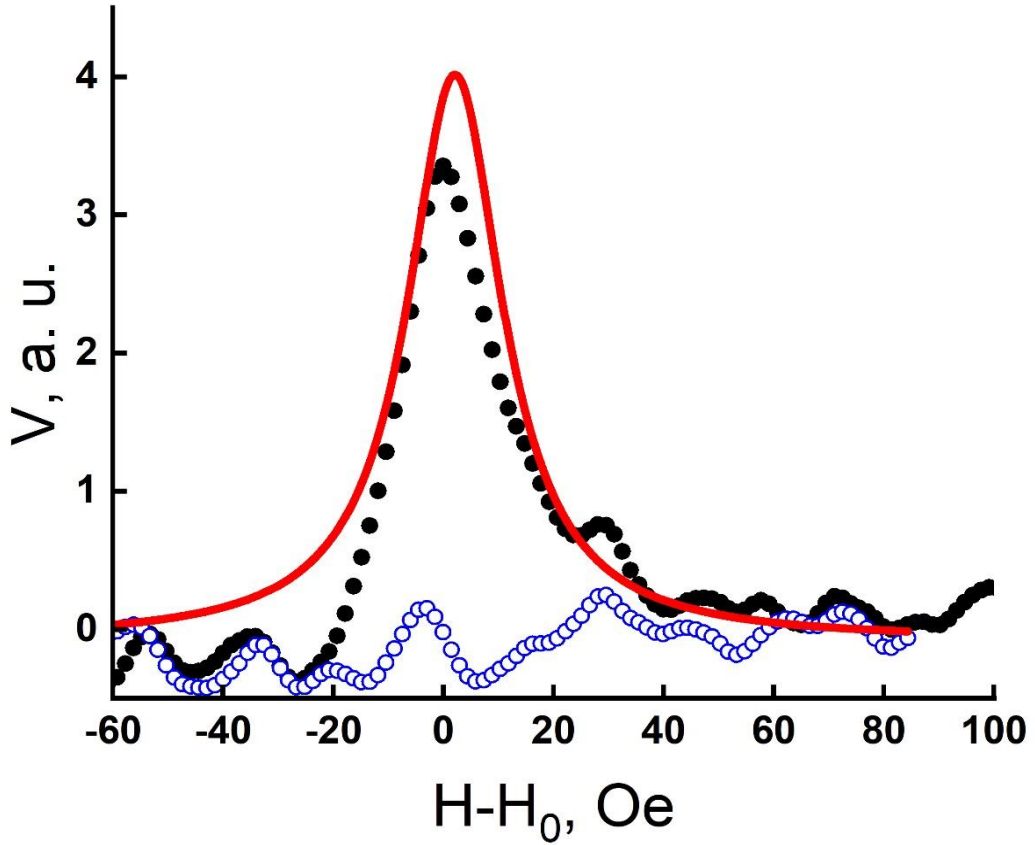

Fig. 3. The voltage response caused by spin current at a frequency of 2.3 GHz and T=300 K. This response is obtained from the half difference of peaks at opposite directions of the dc magnetic field shown in Figure 2. The shape of the spin response is described by the Lorenz function. The contributions from microwave detection by Pt contacts and AMR are an order of magnitude smaller than the amplitude of the spin voltage response.

The flow of spin current through the interface leads an additional damping of spin precession, which is manifested as broadening of the FMR spectrum line $\Delta H$. The broadening is usually determined by the Hilbert spin damping coefficient $\alpha$ [4, 11, 22]. The parameters $\alpha$ and $\Delta H$ are related by the equation [23]: $\Delta H(f) = 4\pi\alpha f/\gamma + \Delta H_0$, where $\gamma$ is the gyromagnetic ratio and $\Delta H_0$ represents the broadening caused by magnetic inhomogeneity. In this analysis, contributions from other damping sources are neglected (see, for example, [24]). The frequency independent broadening $\Delta H_0 = 6 \pm 1$ Oe is small and arises from the magnetic inhomogeneity of the LSMO film in the heterostructure. The values of $\alpha$ determined for the LSMO film in the heterostructure

SIO3/LSMO are $\alpha_{LSMO}$=2.0±0.2 $10^{-4}$ and $\alpha_{SIO/LSMO}$=6.7±0.8 $10^{-4}$ respectively. The increase in Hilbert damping after deposition of SIO3 film allows for an estimate of the real part of the spin mixing conductance Re $g^{\uparrow\downarrow}$ Re $g^{\uparrow\downarrow}$ =(3.5±0.5)·$10^{18}$ m$^{-2}$[4, 21, 25]. Notably, this value is of the same order magnitude as the value of Re $g^{\uparrow\downarrow}$ =1.3 $10^{18}$ m$^{-2}$, determined in [19]. Changing the thickness of SrIrO$_3$ thin film in the heterostructure from 1.5 nm to 12 nm results in a change in the value of Re $g^{\uparrow\downarrow}$ from 0.5 $10^{19}$ m$^{-2}$ to 3.6 $10^{19}$ m$^{-2}$ [18].

According to the theory based on spin interaction between localized and conducting electrons, the spin mixing conductance Re $g^{\uparrow\downarrow}$ is determined by the resistivity $\rho_{SIO}$ and spin diffusion length $\lambda_{SIO}$ of the normal metal with SOI, in this case SIO3 film [25]:

$$\mathrm{Re}g^{\uparrow\downarrow} \approx (h/e^2)/(\rho_{SIO}\lambda_{SIO}). \qquad (3)$$

By employing the equation (3) with $h/e^2 \approx 25.8\ k\Omega$, $\lambda_{SIO}$ = 1 nm [19] and $\rho_{SIO}$ = 3 $10^{-4} \Omega$ cm [12], we obtain $Reg^{\uparrow\downarrow} \approx 8.6 \times 10^{18}$ m$^{-2}$. The obtained value of Re$g^{\uparrow\downarrow}$ is consistent in order of magnitude with experimental data for 3d transition metals, and metallic ferromagnets such as Co, Ni, Fe, which fall in the range of 6 $10^{18}$ - 8 $10^{20}$ m$^{-2}$ [25, 26]. It is important to note that relation (3) is a qualitative estimate of Re$g^{\uparrow\downarrow}$ and does not consider the influence of spin-orbit interaction.

Assuming that the deviation of the $H_0(f)$ dependence for the SIO3/LSMO heterostructure from $H_0(f)$ of the LSMO film can be explained by a change in gyromagnetic ratio $\gamma$, as well as the presence of an imaginary part Im $g^{\uparrow\downarrow}$ of the spin mixing conductance, we calculate a value for Im $g^{\uparrow\downarrow}$ that is significantly higher than what have been estimated for ferromagnetic structures with Pt [4, 22].

When considering more realistic experimental conditions, we believe, that a value of Im $g^{\uparrow\downarrow} \approx 10^{19}$ m$^{-2}$ is more appropriate, taking into account the error of $H_0(f)$ function measurement. This value Im $g^{\uparrow\downarrow}$, as demonstrated in previous studies [12, 22], is comparable to the value of Re $g^{\uparrow\downarrow}$. Furthermore, measurements of the Hall magnetoresistance in Pt/EuS [27] and W/EuO [28] structures indicate that Im $g^{\uparrow\downarrow}$ exceeds Re $g^{\uparrow\downarrow}$ by a factor of 3 and 10, respectively. It is worth noting that the presence of magnetization in the direction perpendicular to the substrate plane could play a significant role, as observed in the case of superlattice composed of SIO3/LSMO heterostructures [17].

Fig. 4 presents the temperature dependence of the amplitude of spin current and the response linewidth of the heterostructure. The spin current amplitude was obtained by dividing the response voltage by the resistance of the heterostructure shown in the inset to Fig. 4. It can be observed that the increase in spin current observed with decreasing temperature in the Pt/LSMO heterostructure [8] is not observed in our case. This discrepancy may be attributed to the presence of a conductive layer at the SIO3/LSMO interface [13].

**3.2 Spin current detection using inverse spin Hall effect.**

The investigation of the inverse spin Hall effect (ISHE) plays a crucial role in the detection of the spin currents [1, 5]. A dimensionless parameter known as the spin Hall angle $\theta_{SH}$ (2) [2, 6] determines the ratio between the spin and charge currents.

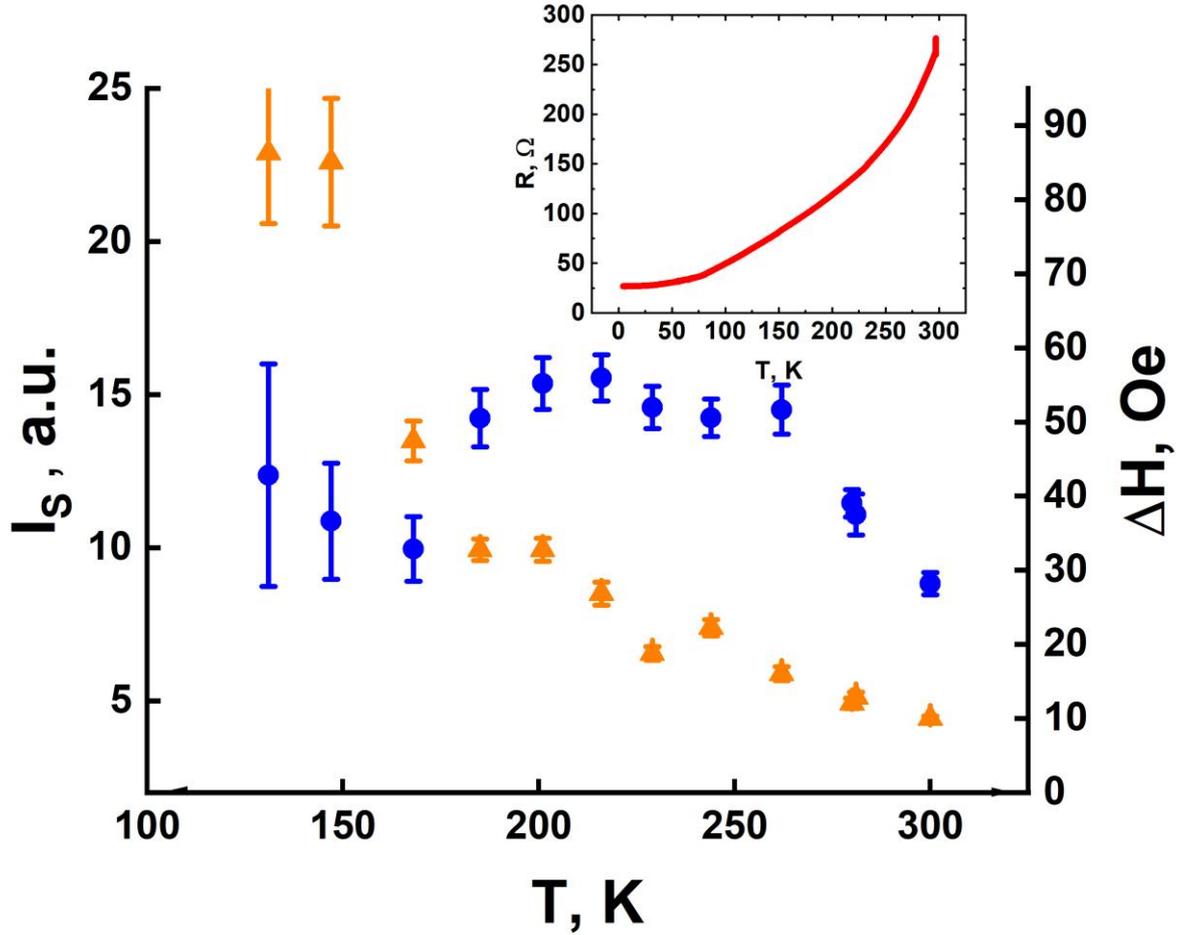

Fig. 4. Temperature dependences of spin current amplitude $I_S$ (filled circles) and line-widths $\Delta H$ (filled triangles). The spin current amplitude is obtained by dividing the voltage response by the resistance of heterostructure. Inset: temperature dependence of heterostructure's resistance at H=0.

To determine the value of $\theta_{SH}$, a Hall geometry arrangement was employed using the SIO3/LSMO heterostructure shown in Fig. 5, along with a 4-point measurement scheme. A dc magnetic field $H$ was applied in the plane of the SIO3/LSMO interface. Proportional voltages $V_L$ (longitudinal magnetoresistance) and $V_T$ (transverse or planar Hall magnetoresistance) were measured. The X direction (see Fig. 5) experienced a current $I=0.5$ mA at a frequency $F=1.1$ kHz and a lock-in amplifier with high sensitivity was utilized for the voltage measurement. By rotating the substrate with the sample around the normal to the substrate, the angle $\varphi$ between the dc magnetic field $H$ and the current $I$ could be varied. The longitudinal magnetoresistance was determined as $R_L=V_L/I$, while the transverse magnetoresistance was determined as $R_T=V_T/I$.

Magnetic-field dependences in normalized units of the change in magnetoresistance $r_{L(T)}=\Delta R_{L(T)}/R_0$ (longitudinal $\Delta R_L$, and transverse $\Delta R_T$) of SIO3/LSMO heterostructure, where $\Delta R_{L(T)}=R_{L(T)}-R_0$ ($R_0$ - magnetoresistance at $H=0$) for fixed angle $\varphi$ between dc magnetic field $H$ and current $I$ were recorded. The obtained values were compared with magnetoresistance measurements for LSMO films, as well as for structures with a platinum film deposited on top of an epitaxial LSMO film (Pt/LSMO). The measured longitudinal magnetoresistance $r_L(\varphi)$ contains in addition to the spin longitudinal magnetoresistance $r_{LS}$ also a contribution from the AMR of the ferromagnetic LSMO film $r_A=R_A/R_0$. Similarly, the transverse magnetoresistance $r_T(\varphi)$ contains also a component of the planar Hall effect magnetoresistance.

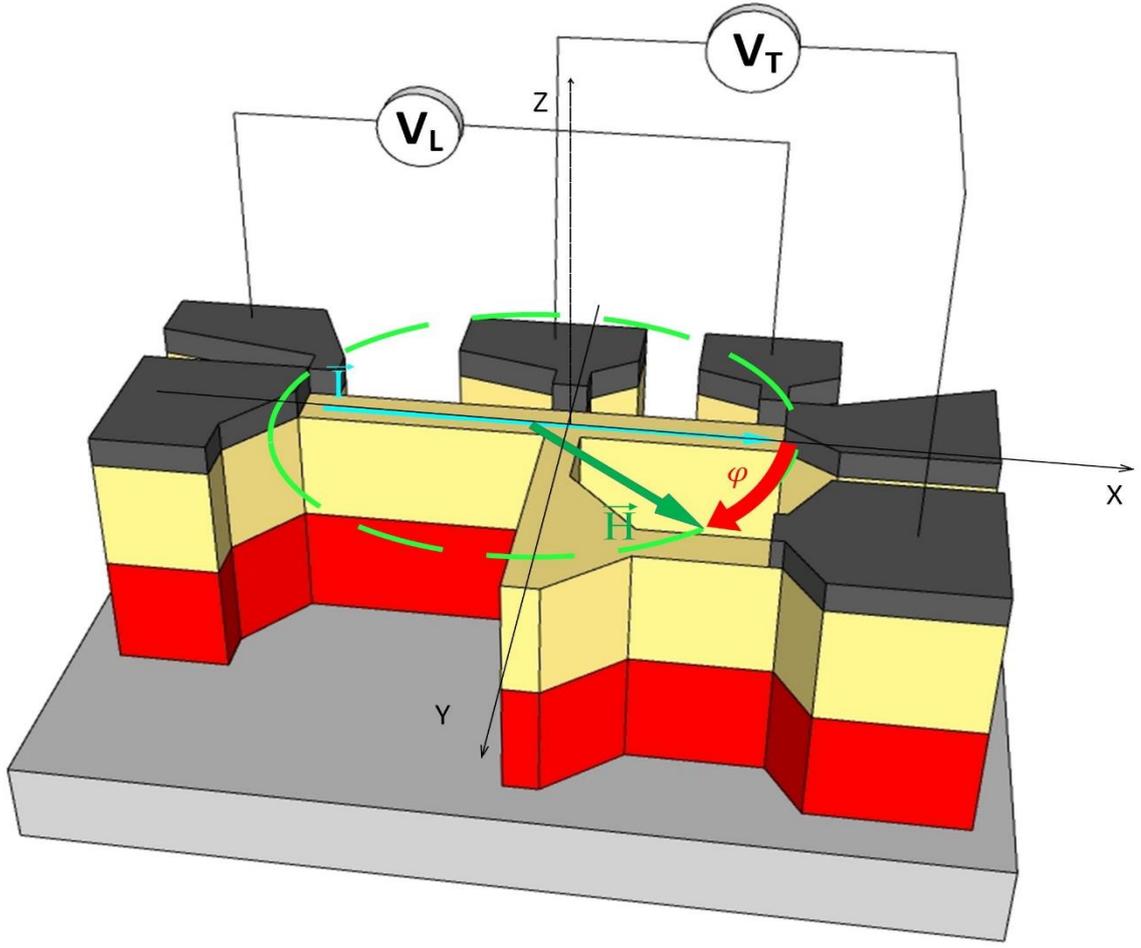

Figure 5. Schematic 3D image of SrIrO$_3$/La$_{0.7}$Sr$_{0.3}$MnO$_3$ heterostructure on (110)NdGaO$_3$ substrate with Pt contact pads. The current I is set along the *X* axis, the angle φ between the dc magnetic field *H* and the current *I* was varied by rotation the sample in the *X-Y* plane.

Fig. 6a illustrates the angular dependence of the longitudinal magnetoresistance $r_L(\varphi)$ in polar coordinates for the SIO3/LSMO heterostructure. The experimental observations indicate that $r_L(\varphi)$ is a parallel connection of $r_{LS}$ and $r_A$. Specifically, the angular dependence of the AMR LSMO film can be described by the function $r_A \cos 2\varphi$, which resembles the dependence of spin magnetoresistance of the SIO3/LSMO heterostructure [6]:

$$r_{LS} = r_1 \cos^2 \varphi \tag{4}$$

where

$$r_1 = \theta_{SH}^2 \frac{\lambda_{SIO}}{d_{SIO}} \operatorname{Re} \frac{2\lambda_{SIO}\rho_{SIO}(\operatorname{Re}G^{\uparrow\downarrow} + i\operatorname{Im}G^{\uparrow\downarrow})}{1 + 2\lambda_{SIO}\rho_{SIO}(\operatorname{Re}G^{\uparrow\downarrow} + i\operatorname{Im}G^{\uparrow\downarrow})} \tag{5}$$

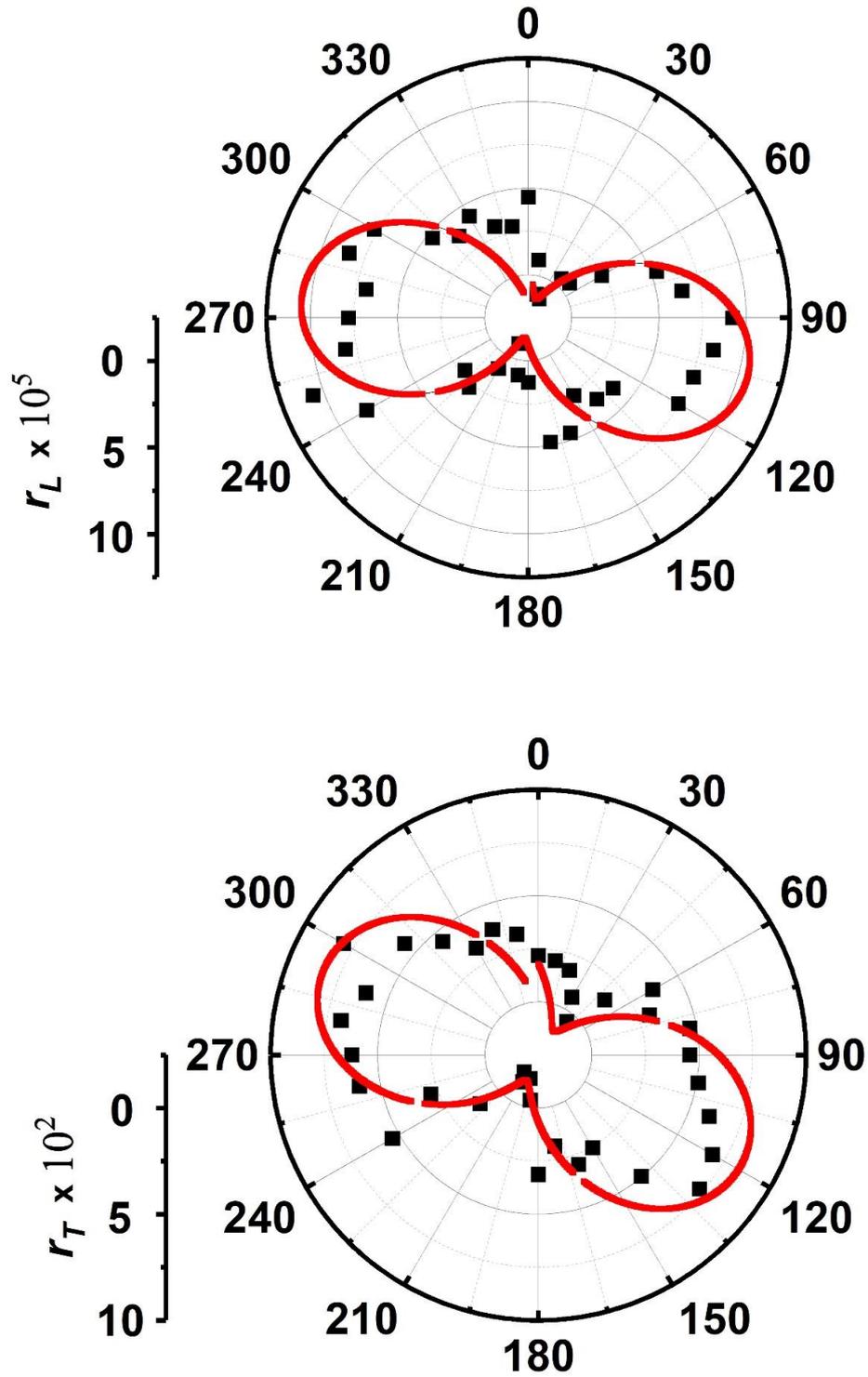

Fig. 6. Angular dependences of normalized values of magnetoresistance of heterostructure $r_{L(T)}$ (squares) and approximation by sinusoidal dependence (solid line) in polar coordinates, taken at the field $H$=100 Oe at $T$=300 K. (a) transverse magnetoresistance, (b) longitudinal. The scale of magnetoresistance variation is shown on the left.

It should be noted that in equation (5) $\text{Re}G^{\uparrow\downarrow}=\text{Re}g^{\uparrow\downarrow}e^2/h$ and $\text{Im}G^{\uparrow\downarrow}=\text{Im}g^{\uparrow\downarrow}e^2/h$ while assuming that the spin diffusion length $\lambda_{SIO}$ is much smaller than the thickness $d_{SIO}$ of the SIO3 film . When a current $I$ flows longitudinally (along the X direction) we obtain a sinusoidal dependence $r_L(\varphi)$ (Fig. 6a). The phase shift in $r_L(\varphi)$ is caused by the difference between the direction of the substrate edge, from which the angle $\varphi$ is measured, and the direction of the LSMO film magnetization easy axis, determined by the crystallographic direction of the [001]NdGaO$_3$ substrate [12]. By utilizing the SIO3 film resistivity $\rho_{SIO}$=3 10$^{-4}$ Ω cm [12], $\lambda_{SIO}$=1 nm [19] and obtained at room temperature and the data for Re$g^{\uparrow\downarrow}$ и Im$g^{\uparrow\downarrow}$ from the Part 3.1. of this paper, we can calculate the spin-Hall angle $\theta_{SH}$ using equations (4) and (5) for thicknesses $d_{SIO}$ =10 nm and $d_{LSMO}$ =30 nm from amplitudes of magnetoresistance $r_L(\varphi)$. The obtained value of $\theta_{SH}$ is approximately 0.03±0.01. This value is an order of magnitude smaller than for the transverse case (Fig. 6b), but around 4 times larger than the spin Hall angle for heterostructures with Pt [3, 6, 8].

Fig. 6b shows the angular dependence of the transverse magnetoresistance $r_T(\varphi)$ in polar coordinates of the SIO3/LSMO heterostructure which, in the general, is the sum of the spin-Hall magnetoresistance $r_{TS}$ [6]:.

$$r_{TS} = \frac{r_1}{2} \sin 2\varphi + r_2 \cos \theta \qquad (6)$$

and planar Hall magnitoresistance of LSMO film $r_H = r_A \sin 2\varphi$. It is important to note that the transverse magnetoresistance also includes a contribution from the out-of-plane magnetoresistance $r_2 \cos \theta$ ($\theta$ is the angle between the current and magnetization along the Z axis, perpendicular to the substrate plane, not shown in Fig. 5). The obtained values of $r_T$ are approximately two order of magnitude larger than in the case of the longitudinal magnetoresistance $r_L$, even when considering only the first term in (6) at $\theta=\pi/2$. Consequently, from the data of $r_T(\varphi)$, a value of $\theta_{SH}$=0.35±0.05 is obtained for the SIO3/LSMO heterostructure. Therefore, the transverse magnetoresistance measurements yield a value of $\theta_{SH}$ about 10 times larger than that from the longitudinal magnetoresistance, where the shunting of the longitudinal AMR magnetoresistance of the LSMO film is available [13]. It should be noted that other methods have also yielded a value of for SIO3/LSMO heterostructures, $\theta_{SH} \cong$ 0.3[11, 16]. The second term in (6), dependent on the imaginary part of the complex spin mixing conductance [6], arises due to magnetization perpendicular the substrate plane and can contribute to an increase in magnetoresistance, as observed in SIO3/LSMO superlattices [17]. Higher values of the spin Hall angle have been reported in structures with SIO3 films: $\theta_{SH}$ =0.76 for Py/SrIrO$_3$ [15] and $\theta_{SH}$ =1.1 for SrIrO$_3$/CoTb$_{1-xx}$ [29]. These values are roughly equivalent to those observed in structures with topological insulators [30].

When the SIO3/LSMO heterostructure was cooled below room temperature T=77 K, the magnetoresistance value decreased. Fig. 7 represents the temperature dependence of the normalized transverse value of $r_T$ for two cases of $r_T(\varphi)$ at fixed values of the angle $\varphi$: when $r_T$ is maximal ($\varphi$ =210º) and minimal ($\varphi$ =275º) at T=300 K. At temperatures bellow150 K the measurement error (not shown in Fig. 7) prevented reliable data extraction. At T=77 K, neither the transverse nor the longitudinal magnetoresistance could be detected. The temperature variation of the transverse resistivity of the SIO3/LSMO heterostructure $R_T$, measured at H=0 is shown in the inset to Fig. 7. Generally, the temperature dependence of $R_T(T)$ exhibits similar characteristics to that of the planar Hall resistantce $r_T(T)$. It is known that the magnetization of an LSMO film increases as the temperature decreases, but equations (4)–(6) do not suggest a change in the spin Hall magnetoresistance with temperature. It should be noted that the temperature dependences of magnetoresistance characteristics, spin diffusion length, and spin Hall angle have been studied in [30, 31] for different structures than those examined in this study. Additionally, it should be noted that the observed increase in spin current with decreasing temperature for Pt/LSMO [8] is not observed in our case, possibly due to the effects of the conducting layer at the SIO3/LSMO interface [12]. Lastly, it should be mentioned that changes in magnetoresistance with temperature could be attributed to the temperature dependences of $Re g^{\uparrow\downarrow}$ and $Im g^{\uparrow\downarrow}$. Modifying the ratio $Im g^{\uparrow\downarrow}/Re g^{\uparrow\downarrow}$ in (5) does not significantly alter parameter $r_1$, but can impact parameter $r_2$ in the presence of out-of-plane magnetization in F-layer.

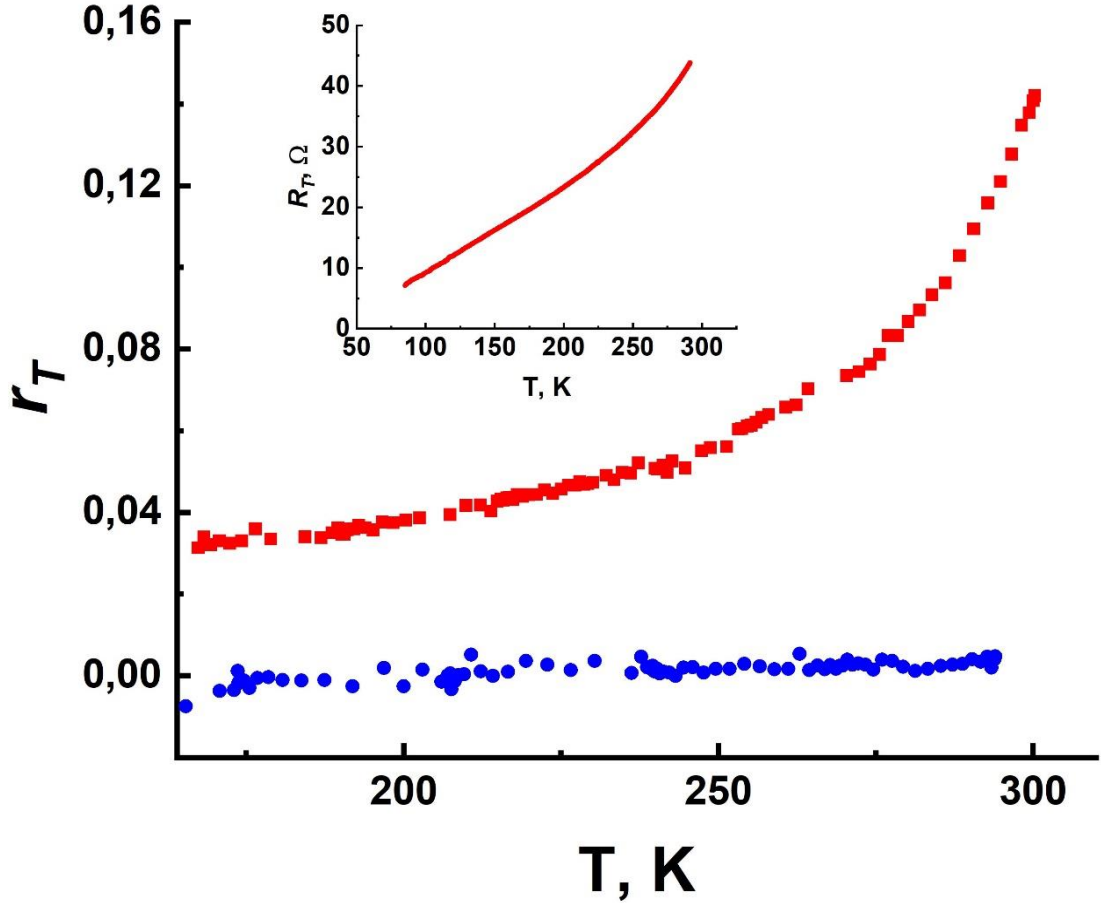

Fig.7. Temperature dependence of transverse magnetoresistance $r_T$. Red squires corresponds to the angle $\varphi=210º$, at which $r_T$ is maximum, blue circles was taken at the minimum value of $r_T$ ($\varphi=275º$). The inset shows the temperature dependence of the transverse resistance at $H=0$

4. **Conclusion**.

The experimental results obtained in this study demonstrate that the spin current response in the $SrIrO_3/La_{0.7}Sr_{0.3}MnO_3$ heterostructure under applied microwaves surpasses the contributions from Pt contacts' microwave detection and anisotropic magnetoresistance by an order of magnitude. Additionally, the real part of the spin mixing conductance at the interface, estimated from the increase in the Hilbert spin damping coefficient, corresponds well with the theoretical estimation based on the spin interaction between localized and conducting electrons. Furthermore, the imaginary part of the spin mixing conductance in the $SrIrO_3/La_{0.7}Sr_{0.3}MnO_3$ interface is found to be an order of magnitude similar to its real part.

Moreover, it is noteworthy that the amplitude of the transverse magnetoresistance angular dependence significantly exceeds that of the longitudinal magnetoresistance. This observation is likely influenced by the shunting effect of the anisotropic magnetoresistance of the $La_{0.7}Sr_{0.3}MnO_3$ film and the conducting layer at the interface of the heterostructure, $SrIrO_3/La_{0.7}Sr_{0.3}MnO_3$. Nevertheless, as the temperature decreases below room temperature, no significant increase in the spin current value was observed either from direct detection of spin response or from the temperature dependence of the spin Hall magnetoresistance.

The authors are grateful to Y.V. Kislinsky, A.A. Klimov, A.M. Petrzhik, and T.A. Shaikhulov for help in the experiment and useful discussions.

The study was funded by a grant from the Russian Science Foundation (project No. 23-49-10006).